\documentclass[preprint2]{aastex}
\usepackage{natbib}
\usepackage{amsmath}
\usepackage{mathrsfs}
\newcommand{\thco}{$^{13}$CO}
\newcommand{\ceo}{C$^{18}$O}
\newcommand{\nht}{NH$_3$}
\newcommand{\cts}{C$_2$S}
\newcommand{\ctfs}{C$^{34}$S}
\newcommand{\nthp}{N$_2$H$^+$}
\newcommand{\htwo}{H$_2$}
\newcommand{\kms}{km\,s$^{-1}$}
\newcommand{\msun}{$M_{\odot}$}
\newcommand{\eg}{{\it e.g.}}

\newcommand{\jypb}{Jy\,beam$^{-1}$}
\newcommand{\core}{L1551-MC}
\slugcomment{Submitted to the Astrophysical Journal}
\shorttitle{A Pre-Protostellar Core in L1551}
\shortauthors{Swift, Welch \& Di Francesco}
\begin{document}
\title{A Pre-Protostellar Core in L1551}
\author{Jonathan J. Swift\altaffilmark{1} and William
  J. Welch} 
\affil{Department of Astronomy and Radio Astronomy Laboratory,
University of California, 601 Campbell Hall, Berkeley, CA
94720-3411}

\author{James Di Francesco}
\affil{National Research Council of Canada, Herzberg Institute of
  Astrophysics, 5071 West Saanich Road, Victoria, BC V9E 2E7, Canada}
\altaffiltext{1}{\bf {\tt js@astro.berkeley.edu}}

\begin{abstract}
Large field surveys of {\nht}, {\cts}, {\thco} and {\ceo} in the L1551
dark cloud have revealed a prolate, pre-protostellar molecular core
({\core}) in a relatively quiescent region to the northwest of the
well-known IRS 5 source. The kinetic temperature is measured to be
9\,K, the total mass is $\sim 2\,M_{\odot}$, and the average particle
density is $10^4$--$10^5$\,cm$^{-3}$ . {\core} is $2^\prime\!\!.25
\times 1^\prime\!\!.11$ in projection oriented at a position angle of
133$^\circ$. The turbulent motions are on the order of the sound speed
in the medium and contain 4\% of the gravitational energy, $E_{grav}$,
of the core. The angular momentum vector is projected along the major
axis of {\core} corresponding to a 
rotational energy of $2.5 \times 10^{-3}\,\sin^{-2}i\,|E_{grav}|$. The 
thermal energy constitutes about a third of $|E_{grav}|$ and the
virial mass is approximately equal to the total mass. {\core} is
gravitationally bound and in the absence of strong, $\sim 160\,\mu$G,
magnetic fields will likely contract on a $\sim 0.3$\,Myr time
scale. The line profiles of many molecular species suggest that the
cold quiescent interior is surrounded by a dynamic, perhaps infalling
envelope which is embedded within the ambient molecular gas of L1551.
\end{abstract}


\section{Introduction}
Stars play a vital role in the cosmos, enriching the interstellar
medium and providing the basic building blocks for the formation of
planets and life. Stars with masses $\lesssim 1${\msun} are vastly
more numerous than higher mass stars and contain most of the stellar
mass in our Galaxy. Much advancement has been made toward the
understanding of low mass star formation due in part to the study of
nearby star-forming molecular complexes such as Taurus-Auriga. The
dense molecular environments most closely associated with the star
formation process have been identified~\citep{bei86,mye87} and
characterized in large survey data~\citep[][and references
  therein]{jij99}. However, the initial conditions under which
protostellar collapse occurs remain poorly understood. 

Dicarbon monosulphide is a powerful probe of the dense interstellar
medium in star-forming regions containing a multitude of emission
lines in the millimeter and centimeter wavebands with no hyperfine
structure~\citep{sai87}, a large Zeeman splitting factor~\citep{shi00}
and high critical density~\citep{wol97}. Observational and theoretical
studies of {\cts} in the dense interstellar medium suggest that it
traces the early evolutionary stages of pre-protostellar
cores~\citep{suz92,lee03} and is otherwise not detectable in star
forming regions~\citep{lai00}.

In 2001, we conducted a {\cts} survey of Lynds
1551~\citep{lyn62,sne81} in search of environments in the early stages
of star formation. L1551 is an $\sim 80\,${\msun} dark cloud
at a distance of 140\,pc~\citep{ken94} associated with more
than 20 known pre-main sequence stars~\citep[see][and references
  therein]{pal02}. The energetics from the embedded young stellar
objects are witnessed as molecular outflows~\citep{sne80,mor95},
jets~\citep{mun90} and numerous Herbig-Haro objects~\citep{dev99}.

A single emission region appeared in the northwest quadrant of our
{\cts} survey which we shall refer to as {\core} (Lynds 1551 Molecular
Core). While {\core} is well separated from the known energetics in
L1551, there is a low excitation Herbig-Haro object, HH265, which lies
at the edge of the projected {\cts} emission~\citep{dev99,gar92}. No
known energetics in L1551 are directed toward this region (see above
references and also Figure~\ref{overview}) and the driving source for
HH265 remains a mystery.

The many observations at centimeter and millimeter wavelengths used to
investigate {\core} are described in~\S\ref{sobs} from which the
morphological and physical properties are derived in
\S\ref{sresults}. A discussion of these results is given in
\S\ref{sdisc}, and the conclusions, including future study plans, are
summarized in~\S\ref{sconclusions}.

\section{Observations and Reductions} \label{sobs}
Table~\ref{obstable} displays the general parameters of our
observations while the text in this section summarizes the relevant
observational information not included in the Table.

\subsection{BIMA Observations}
\subsubsection{$\lambda = 9$\,mm} \label{9mm}
On 2001 September 2 and 3 a $\sim 25^\prime \times 25^\prime$ region
centered on L1551 was mapped in {\cts} using the
BIMA\footnote{The Berkeley-Illinois-Maryland Association array is
  operated by the University of California, Berkeley, the University
  of Illinois, and the University of Maryland with support from the
  National Science Foundation.} array in its
most compact (D) configuration equipped with 30\,GHz 
receivers~\citep{car96}. A 7 pointing mosaic ($\sim 10^\prime \times
10^\prime$) centered on the peak emission detected in the D array
configuration was then observed in the B and C array configurations in
the fall of 2001 and 2002. Only the C and D array data are used (see
\S\ref{senv}) which have sufficient $u-v$ coverage to sample all
spatial frequencies between 27{\arcsec} and $4^\prime\!.5$. 

These observations utilized the highest resolution correlator
configuration at BIMA~\citep{wel96} which translates to a
0.054\,{\kms} velocity resolution at the {\cts} transition
frequency. Phase calibration was ensured with regular scans of the
quasar 0449+113 while the flux scale of the observations was
determined with scans of Jupiter and 3C84. All data were
reduced with the MIRIAD reduction package~\citep{sau95} using standard
MIRIAD tasks.

\subsubsection{$\lambda = 3$\,mm}
The {\cts} emission region described above ({\core}) was then mapped
at the transition frequencies of a number of molecular tracers in the
3 mm band. These mosaics all consist of a 7 pointing, Nyquist sampled
hexagonal grid centered on $4^{\rm h}31^{\rm m}09^{\rm s}\!.1$,
$+18^\circ12^\prime09^{\prime\prime}\!\!.5\,\,(J2000)$ covering a
$\sim 2^\prime\!.5$ region with 100\% sensitivity.   
These tracks were run at the Hat Creek Radio Observatory in the
C and D array configurations between 2002 April and June. In addition
to the spectral line information, each correlator configuration
contained a total of 800\,MHz of continuum which was calibrated and
preserved. 

The $u-v$ coverage was good at all molecular transition frequencies
giving well sampled maps for sky angles $15^{\prime\prime}\lesssim
\theta \lesssim 100^{\prime\prime}$ except for CS which was only
observed in the C array configuration and is more sparsely sampled
over this spatial range. The integration time per mosaic pointing
totaled $\sim 45$ minutes for the CS map and $\sim 2$ hours for all
others. System temperatures were variable from $\sim 250$ to
650\,K. All tracks contained 5 minute scans of 0530+135 or 0449+113
at intervals of $\lesssim 25$ minutes for phase calibration and 5
minute integrations on Saturn and W3OH for flux calibration. All data
were reduced with the MIRIAD reduction package. 

\subsection{GBT Observations of NH$_3$} 
Single pointing spectra and spectral line maps containing the (1,1)
and (2,2) inversion transition frequencies of the {\nht} molecule were
obtained with the NRAO\footnote{The National Radio Astronomy
  Observatory is a facility of the National Science Foundation,
  operated under cooperative agreement by Associated Universities,
  Inc.} 100~m Robert C. Byrd Green Bank Telescope (GBT) in K-band on
the nights of 2004 January 6th and 10th. The spectrometer was set to
contain 50\,MHz of bandwidth centered on a 23.7088144\,GHz sky
frequency using 9 levels of sampler resolution with 32768 lags in dual
polarization mode. Therefore, both the (1,1) and (2,2) inversion
transition frequencies including the quadrupole satellites fell within
the spectral window with 0.019\,{\kms} spectral resolution. The
absolute reference position was chosen to lie north of L1551 at
$4^{\rm h}31^{\rm m}20^{\rm s}\!$,
$+18^\circ20^\prime00^{\prime\prime}\,\,(J2000)$.

The spectral line maps are the result of the On-the-Fly (OTF) mapping
technique and cover approximately $5^\prime\times5^\prime$ centered on
$4^{\rm h}31^{\rm m}08^{\rm s}\!.2$,
$+18^\circ13^\prime07^{\prime\prime}\!\!.5\,\,(J2000)$. The GBT main
beam efficiency, $\eta_B = 0.81$ and the aperture efficiency, $\eta_A
= 0.58$ in K-band. The 60\,arcsec s$^{-1}$ scan rate and 13{\arcsec}
row spacing used to make the maps sufficiently sample all spatial
frequencies available to the GBT within the mapped region. System 
temperature calibrations performed every two rows minimize gain
variations across the individual maps and alternating the map scan
directions sequentially between right ascension and declination
minimize systematic errors ({\eg} striping). Typical system
temperatures were $\sim 35$\,K on both observing nights.

The single pointing spectra were calibrated and exported into ASCII
tables with the AIPS++ software package. All further reductions were
performed using the IDL software of Research Systems, Inc. The OTF
maps were calibrated and gridded in AIPS++ then written to FITS format
for further processing in IDL.  

Our final spectra, including those in the spectral maps, are the
combination of the two independent, calibrated polarization channels
fit in off-line frequencies with a third-order polynomial. Our final
spectral map is the combination by weighted mean of six individual
maps. Systematic effects from the OTF mapping technique are noticeable
in the channels of the final map, although reduced significantly in
the velocity integrated map. The total flux in the final spectral map
was scaled up by 10\% to match the total flux in our individual
spectra taken at positions within the mapped region.  

\subsection{Kitt Peak 12\,m Observations at $\lambda = 3$\,mm}
\subsubsection{On-the-Fly {\thco} and {\ceo} Maps}
A $20^\prime\times 20^\prime$ region of L1551 was mapped in
{\thco}$(1\rightarrow0)$ on 2000 November 21 and 22 with the Kitt Peak
12~m Telescope\footnote{The Kitt Peak 12~m Telescope is operated by
  the Arizona Radio Observatory at Steward Observatory, University of
  Arizona, with partial support from the Research Corporation.} using
the OTF observing technique. The Millimeter Autocorrelator (MAC) was 
configured in 2 IF mode with 8192 channels and 150\,MHz bandwidth. The
main beam, taken to be a 2-dimensional Gaussian with a full width at
half-maximum (FWHM) of 55\arcsec, comprises 90\% of the full beam
pattern. The aperture efficiency of the 12m at this observing
frequency is 0.5. The 60\,arcsec s$^{-1}$ scan rate and 19{\arcsec} row
spacing of the maps provide spatial over-sampling. The number of rows
per reference scan (OFF) and the number of OFFs per total power
calibration varied over the course of the observations to keep the
system temperature variations below 3\% between measurements. System
temperatures on the 21st were very good, $\sim 180$\,K, and decent on
the 22nd, $\sim 275$\,K.  

A comparable OTF map in {\ceo}$(1\rightarrow0)$ observed on
2002 January 15 and on three consecutive nights starting on 2002
October 21 covered the central $17^\prime \times 17^\prime$ of
L1551. The backend setup was identical to the {\thco} 
observations. System temperatures were good, $\sim 250$\,K for our
January observations and variable for the three nights in October.

The raw OTF data were calibrated, converted to visibility data,
baseline fit using a first order polynomial, gridded into an image and
then written to FITS file format with the AIPS reduction software. 
Further analyses were performed using IDL.

\subsubsection{CS and {\nthp} Spectra}
Single pointing spectra were taken on the night of 2002 November 30 at
selected points within the {\cts} emission region at the transition
frequencies of both {\nthp}$(1\rightarrow0)$ and
CS$(2\rightarrow1)$. The frequency throw was set to $\pm1.0\,$MHz for
CS and $\pm3.0\,$MHz for {\nthp}. The MAC was configured in 2 IF mode
with 12288 usable channels within a 75\,MHz band.  

Each spectrum was calibrated and folded using UNIPOPS at the
observatory and then output into FITS tables. A third order
polynomial baseline fit, the combination of independent polarizations,
and further analyses were all done using IDL.  

\section{Results} \label{sresults}
\subsection{Overview} \label{sover}
Figure~\ref{overview} shows our full {\thco} Kitt Peak 12~m map with
known outflow sources overlaid in the left hand panel. The right
hand panel is a closeup of the localized region of {\cts} emission
L1551 revealed by our BIMA survey. The {\thco} is optically thick in
this region and has a relative maximum which coincides with the peak
of the {\cts} emission. HH265 lies at the eastern edge of the
projected {\cts}.

Figure~\ref{pointsource} shows a prolate, dense core seen in {\nht}
spatially overlapping with the {\cts} emission. There are no
compelling young stellar object candidates near {\core} in either
radio, infrared, or X-ray data. Figure~\ref{nh3plot} reveals the
morphology of the {\nht} emission to more closely resemble the
distribution of {\ceo} than the optically thick {\thco}.

Figure~\ref{spectra} shows a comparison of many molecular spectra
taken at the peak of the {\cts} emission, defined as Point A, relative
to the central velocity of the {\nht} emission. All lines in
Figure~\ref{spectra}, excluding {\cts}, are from the 12~m
observations; BIMA observations of these same transitions did not
result in any detections, indicating the overall emission is spatially
smooth. The {\cts} emission peaks significantly to the red of the
reference velocity. There are signs of self-absorption in both the
{\thco} and CS spectra. The CS line profile in relation to the central
velocity of the {\nht} emission is consistent with an infalling
envelope of {\core}. The {\nthp} profile is slightly asymmetric,
centered somewhat to the red of the {\nht} central velocity, but is
otherwise consistent with the {\nht} emission profile.

\subsection{Is L1551-MC Starless?} \label{sstarless}
Jets, winds and outflows from young stellar sources can dramatically
affect the interstellar environment from which they are
created. Therefore it is important to determine whether or not a
stellar point source is present within a molecular core before the
molecular line data are interpreted.

The emission at all {\it IRAS} wavelengths in L1551 is dominated by
the L1551-IRS5/L1551NE and HL/XZ Tau regions and shows no evidence of
any other sources within L1551. A recent paper by \cite{gal04} reports
the results from a deep mid-IR search for young stellar objects in
L1551 using the Infrared Space Observatory (ISO) and four sources
(sources 20, 23, 29 and 32 of their Table~2) lie within a 3{\arcmin}
radius of {\core}. Source 32 is not considered a potential protostar
candidate since it was only marginally detected in the 14.3\,$\mu$m
band and has neither an optical nor near-IR counterpart. 

The three sources detected at 6.7\,$\mu$m are plotted on an $(H-K)$
vs. $\left(K-m_{6.7}\right)$ color-color diagram~\citep[Figure 4
  in][]{gal04}. Sources 23 and 29 are consistent with dust reddened
stars with no infrared excess. Source 20 was not detected in the $H$
band. However, this lower limit on $m_H$ places it above the reddening
line at a position consistent with being a reddened main-sequence
star.

A SIMBAD search reveals one additional radio source within the 3{\arcmin}
radius of the peak {\nht} emission, GRL11. This 1.4\,GHz continuum
source is most likely of extragalactic origin~\citep{gio00,sne86}. An
X-ray source, BFR12, lies within 4{\arcsec} of GRL11 and is likely
related~\citep{bally03}. An additional X-ray source from the Chandra
survey of L1551, BFR2, is seen to the southwest of {\core} and is
also most likely a background object.

The combination of all our 3 mm BIMA continuum data precludes the
existence of a compact, $< 10$\arcsec, source at
$\sim 100$\,GHz brighter than 5.3\,mJy with 99\% confidence. Given these
considerations we conclude that {\core} is starless. A plot of
the~\cite{gal04}, 2MASS and SIMBAD sources can be seen in
Figure~\ref{pointsource}.

\subsection{Morphology} \label{smorph}
\subsubsection{Distribution and Orientation of Dense Gas} \label{sdense}
The dense, inner regions of cores are traced well with {\nht} emission
due to a high abundance under these conditions and a tendency to avoid
depletion onto grains~\citep{aik03}. In this section the primary
morphological features of {\core} are characterized based on the
{\nht} emission in the region. Several single pointing spectra 
taken at 23\,GHz within the larger L1551 cloud reveal that {\nht}
emission is not widespread but rather confined to the regions around
L1551NE and IRS5 in addition to the region mapped in this paper (see
Figure~\ref{nh3plot}). The results from these spectra
are summarized in Table~\ref{nh3pnts}.

{\core} appears roughly elliptical in projection with an axial ratio,
$y=2$. Figure~\ref{moment1} shows the velocity integrated {\nht}(1,1)
emission in the optically thin satellite components as contours. This
emission distribution is fit with a 2-dimensional 
Gaussian with the peak emission amplitude, 2-d centroid, major
and minor widths, and position angle as free parameters. The gross
morphological features, including the characteristic quantities from
the average spectrum within a half-maximum {\nht} contour are summarized
in Table~\ref{morphtable}.

A first moment map is created by fitting each spectrum in our
{\nht} data cube according to \S\ref{snht}. Local velocity gradients
calculated at each pixel by vector sum of the central velocity
difference at the eight neighboring pixels in the first moment map are
shown in Figure~\ref{moment1}. There is some scatter to the local
gradients, but there is an overall trend of higher velocities toward
the southwest. Fitting a velocity gradient to the map in a manner
similar to that done by \cite{goo93} yields $\mathscr{G} = 1.2 \pm
0.1$\,{\kms}pc$^{-1}$ at a position angle of $224\pm 5^\circ$, aligned
along the short axis of the velocity integrated emission. The vector
addition of all local velocity gradients in Figure~\ref{moment1}
yields an identical gradient position angle but smaller
amplitude. These results suggest that {\core} is prolate. (We assume a
prolate spheroidal morphology for {\core} in forthcoming calculations
unless otherwise stated.) 

The velocity integrated emission in all satellite components averaged
over elliptical annuli with an axial ratio of two centered on the
$(\alpha,\delta)$ coordinates in Table~\ref{morphtable} at 13{\arcsec}
radius intervals is shown in Figure~\ref{profile}. The 2-dimensional
projected radial distance from center is defined as $b$, the impact
parameter. The simplest characterization of the intensity profile is
of a power law form, $I(b) \propto b^{-p}$. If this function is
assumed to describe the intrinsic shape of the intensity profile it
becomes immediately apparent that it grossly overestimates the flux in
the central pixels. A two component (``broken'') power law
function of the form 
\begin{equation}
I(b) = 
\begin{cases} 
  c_1 b^{-p_1} & : \quad b < b_0 \\
  c_2 b^{-p_2} & : \quad b > b_0 \\
\end{cases}
\label{brokenpl}
\end{equation}
where $c_2 = c_1 b_0^{(p_2 - p_1)}$, is much more successful in
fitting these data. The fit according to the minimum chi-squared value
of this intrinsic radial profile convolved with the GBT beam is shown
in Figure~\ref{profile} as the solid line. There is little statistical
uncertainty in our best fit parameters, but there is an unknown
systematic error since the true 3-dimensional shape of {\core} cannot
be determined directly from our data. The standard deviation in these
parameters invoked by using elliptical annuli of varying axial ratios
between $y=1$ and $y=2$ gives an estimate of this systematic error. The
best fit parameter values are $p_1 = 0.15\pm 0.04$, $p_2 = 1.48\pm 0.03$,
and $b_0 = 0.040\pm0.005$\,pc. 

An intrinsic intensity profile model of the form 
\begin{equation}
I(b) = \frac{I_0}{1+(b/b_0)^\gamma}
\label{profilefunc}
\end{equation}
fits our data to even higher precision than the broken power law. A
similar model has been used by \cite{taf02} to fit the 3-dimensional
radial profile of molecular cores with great success and is a good
approximation to a Bonner-Ebert profile~\citep{taf04}. This model is
used here to fit the 2-dimensional intensity profile only for its
convenience and easily interpretable parameters. The minimum
chi-squared values for the free parameters of this function are $I_0 =
2.4\pm 0.2$\,K\,{\kms}, $b_0 = 0.05\pm 0.01$\,pc, and $\gamma =
2.2\pm0.1$. This fit is displayed as the dashed line in
Figure~\ref{profile}.

A broad brush picture of the mass distribution in {\core} can
be inferred if it is assumed to be an optically thin prolate spheroid
at constant temperature, constant abundance fraction, and in local
thermodynamic equilibrium (LTE). The inferred flat inner density
profile with power index $\sim 1.2$ and break at $\sim 0.04$\,pc where
the power law index steepens to $\sim 2.5$ is typical for a starless
core in Taurus~\citep[{\eg}][]{cas02}.

This simple picture is most challenged by the possibilities of a
varying {\nht} abundance fraction and a non-symmetric geometry in the
third dimension. Recent studies suggest that the abundance fraction of
{\nht} increases toward the center of cores~\citep{taf04}, so it is
unlikely that an abundance fraction variation alone could produce the
flattened inner profile. It would also be unlikely that a
geometrical effect could reproduce the effect of having a centrally
flat intensity distribution. The existence of a turnover radius at
$b_0 \approx 0.04$\,pc is therefore robust, while the slope of the
profile both inside and outside $b_0$ is not. No significant {\nthp}
emission is detected in the BIMA 3 mm data, sensitive to spatial scales
between 15 and 80{\arcsec} at a 3-$\sigma$ level of 0.1\,{\jypb},
suggesting that the densest gas in {\core} is distributed smoothly.

\subsubsection{The Envelope} \label{senv}
Carbon monoxide, the second most abundant molecule in the Galactic
interstellar medium, has a relatively small dipole moment and traces
low density molecular environments. Figures~\ref{overview} and
\ref{nh3plot} show {\thco} and {\ceo} to be pervasive in L1551 and
therefore these isotopomers probe the low density envelope of
{\core} as well as the environment in which it sits.

If it is assumed that the rotational excitation temperature of {\thco}
and {\ceo} are equal, the optical depths of these isotopomers can be
estimated. The ratio of antenna temperatures of the {\thco} and {\ceo} 
transitions are
\begin{equation}
\frac{T_A^*(^{13}{\rm CO})}{T_A^*({\rm C}^{18}{\rm O})} = 
\frac{(1-e^{-\tau_{13}})\,\eta_{13}}
{(1-e^{-\tau_{18}})\,\eta_{18}}
\label{tauratio}
\end{equation}
where the subscripts 13 and 18 are used to denote quantities related
to {\thco} and {\ceo} respectively and $\eta$ is the beam
efficiency. The telescope parameters are nearly
equivalent at these neighboring transition frequencies so to a good
approximation $\eta_{13} = \eta_{18}$. It has been shown that for high
column molecular gas ($A_V > 3.5$\,mag) the {\thco}/{\ceo} abundance
ratio is relatively uniform and 
consistent with terrestrial abundance ratios~\citep{mcc80} which
implies $\tau_{13} = 5.4\,\tau_{18}$. Using these numbers, the
mean optical depth of {\thco} in the {\nht} emission
region is 1.9. 

The BIMA {\thco} and {\ceo} maps show a negligible amount of total
emission. There is low level emission visible in the line wings of
our {\thco} map between the LSR velocities of
5.9--6.5\,{\kms} and 7.2--7.9\,{\kms}. However, there is not enough
signal above the 0.44\,K per channel noise for any interpretation. No
emission is detected in the line core where $\tau_{13} > 1$. The
peak antenna temperature in our Kitt Peak maps are 5.74\,K and 2.23\,K
in {\thco} and {\ceo} respectively and our BIMA maps are well sampled
between 15{\arcsec} and 80{\arcsec} with an RMS of $\sim
0.4$\,K. Therefore, the {\thco} and {\ceo} are tracing a relatively
smooth envelope of {\core}, with an insignificant amount of structure
smaller than 80{\arcsec}.

The {\thco} and {\ceo} emission extends smoothly from the region
around {\core} into the greater L1551 cloud. The
optically thin {\ceo} follows more closely the overall form of {\nht}
emission showing no correlation to the {\cts} map. Therefore it
appears that most of the mass in the envelope follows the gross
morphology of the dense inner regions characterized in
\S\ref{sdense}. This interpretation, however, does not account for
depletion of these molecular species onto grains, which is known to
occur in dense cores~\citep{aik03,lee03,taf04}.

The {\cts} emission in this region curls around the southwestern limb
of the {\nht} emission and stretches to the location of HH265 (see
Figure~\ref{pointsource}). The brightest peak of {\cts} emission,
defined as Point A in this paper, coincides with a relative maximum in
the optically thick {\thco} emission (see Figure~\ref{overview}) but
shows no such correlation in the optically thin {\ceo}.

No {\cts} emission is seen to a 3-$\sigma$ level of 0.27\,{\jypb}
using BIMA in its B array configuration which samples spatial
frequencies primarily between 10{\arcsec} and 35{\arcsec} at the our
observing frequency. These data
are not used in our final maps and spectra, yet the presented maps
reflect an accurate morphology of this tracer given our sensitivity
limits.

The CS spectrum at Point A is highly asymmetric containing a blue peak
and a red shoulder, consistent with self-absorption in an infalling
layer or shell~\citep{mye96,eva99}. The {\cts} emission is
significantly redshifted with respect to the systemic {\nht}
velocity. If the {\cts} emission region lies on the near side of the
core, the arcuate distribution of {\cts} may represent an infalling
layer around {\core}. The possibility of infall in {\core} is
discussed further in \S\ref{senvi}.

\subsection{Physical Properties} \label{sphys}
Many useful quantities describing conditions within dense molecular
environments can be derived from the {\nht}(1,1) and (2,2) inversion
transitions~\citep[see][]{ho83}. Our {\nht}(1,1) spectra are
fit according to the method described in \S\ref{snht}. Minimum
chi-squared values less than unity, as expected for a detailed model,
were typically achieved using this method and suggest that the LTE
assumption is valid for this tracer. Table~\ref{nh3qtable} summarizes
the results of our {\nht} fitting and Figure~\ref{pointaspec} shows
fits to the spectra at Point A.

\subsubsection{Kinetic Temperature} \label{sTk}
The flux ratio of the {\nht} (1,1) and (2,2) transitions can be used
to estimate the kinetic temperature of the emitting molecular gas (see
\S\ref{snht}). The best fit parameters for the (1,1) and (2,2)
transitions along with Equations~\ref{trot} and \ref{tk} give a
kinetic temperature at Point A, $T_k = 8.7 \pm 0.7$\,K.

The sensitivity of our {\nht} OTF map was not sufficient to detect the
(2,2) transition. While a variation in gas temperature by a factor of
$\sim 2$ has been conjectured for Taurus cores~\citep{gal02}, the
upper limit on $T_k$ from the Point B spectrum limits the temperature
variation to $< 50$\% (see Table~\ref{nh3qtable}). A constant kinetic
temperature $T_k = 9 \pm 1$\,K within {\core} is assumed in all
following calculations which include temperature.

\subsubsection{Thermal and Turbulent Motions} \label{svelocity}
The full width at half-maximum thermal line width for a gas in LTE at 
kinetic temperature $T_k$ is given by
\begin{equation}
\Delta v_{th}^2 = 8\ln2\frac{kT_k}{m}
\label{vth}
\end{equation}
where $m = 2.33$\,$m_H$ is the mean particle mass of a gas of
molecular hydrogen containing atomic helium with a solar mass fraction
($Y=0.28$). Using $T_k$ from \S\ref{sTk}, $\Delta v_{th} =
0.42$\,{\kms}. The non-thermal line width of a molecular species can
be derived from the observed line width, $\Delta v_{obs}$, and kinetic
temperature.
\begin{equation} \Delta v_{nt}^2 = \Delta v_{obs}^2 -
  8\ln2\frac{kT_k}{m_{mol}}
\label{vnt}
\end{equation}
where $m_{mol}$ is the mass of the observed species. Assuming well
mixed conditions, this quantity represents the non-thermal line width
for the medium. The total line width can be obtained by adding the
thermal and non-thermal components in quadrature.

The mean non-thermal line width is $\Delta v_{nt} = 0.20 \pm
0.002$\,{\kms} and the mean total line width is $\Delta v_{tot} = 0.46
\pm 0.01$\,{\kms} obtained by averaging the line widths from each
spectrum fit according to \S\ref{snht} within a half-maximum {\nht}
contour. The errors in the line widths are from 
\cite{lan82} and by propagating the error in $T_k$. The sound speed in
the medium is given by, $(k\,T_k/m)^{1/2} = 0.19\pm0.01$\,{\kms},
roughly equal to $\Delta v_{nt}$ in the region. There is no
significant correlation between the projected distance from the peak
{\nht} emission and the velocity width.

\subsubsection{Mass and Density} \label{smass}
The total mass of a molecular core cannot be observed directly and
must be inferred using estimates of tracer abundances. The most
abundant molecule, {\htwo}, does not radiate in the millimeter or
centimeter bands due to its homo-polar geometry, leaving mass estimates
of molecular regions dependent on conversion factors for less abundant
molecules such as CO. Many processes affect molecular abundance ratios
\citep[see][for a comprehensive review]{van98} and published results
for $X$(CO) in Galactic molecular clouds have varied by a factor of
$\sim3$~\citep{lac94}. Following is an estimate of the total mass of
{\core} first using our CO isotopomer data and {\nht} data within a
half-maximum contour of the velocity integrated {\nht} intensity and
then by consideration of the virial theorem.

The total column of {\thco} in each pixel of {\core} is found assuming
that {\thco} is thermalized at 9\,K. These values are corrected for
optical depth using Equation~\ref{tauratio} and then summed. Using
$X$({\thco})$ = 1.12\times10^{-6}$ converts this number to a total
mass of 1.9\,{\msun}.  

The total column of {\nht} in the (1,1) state is derived approximating
the partition function as the sum over all metastable states up to $J
= 6$~\citep{win79}. The total {\nht} mass of {\core} is found to be
$2.4\times10^{-7}$\,{\msun} which implies a total molecular mass
between 0.8--2.4\,{\msun} for $X$({\nht}) between
$4.0\times10^{-8}$~\citep{jij99} and
$1.5\times10^{-8}$~\citep{taf04}. The total mass of {\core} estimated
from molecular line emission is $M_{mol} \sim 2$\,{\msun} good to
within about a factor of 2.

The virial mass is obtained assuming virial equilibrium between
thermal and gravitational self-energy. If it is assumed that {\core}
is comprised of molecular 
hydrogen
\begin{equation}
M_{vir} = \frac{25\,\Delta v^2\,R}{24\,\alpha_G\,G\,\ln2}
\label{mvir}
\end{equation} 
Here $\alpha_G$ is the geometrical and non-uniformity factor of
Equation~\ref{egrav} which is of order unity~\citep{ber92}, $\Delta v
= \Delta v_{tot}$ from Table~\ref{phystable}, and $R$ is the
semi-minor axis of the prolate spheroid. Equation~\ref{mvir} yields
$M_{vir} = 1.9^{+0.2}_{-0.7}${\msun} for $\alpha_G = 0.87$
 
An estimate of the volume density from the {\nht} fits is presented in
Table~\ref{nh3qtable}. This density, $n_{ex}$, is estimated from the
excitation temperature of the fit assuming non-LTE conditions and is a
lower limit to the true density of the medium in the derivation (see
Equations~\ref{nH2} and following). {\nthp}($1-0$) is less likely to
be in LTE than {\nht}(1,1) owing to a much higher critical
density. The hyperfine structure in the {\nthp}($1-0$) transition is
fit in a similar fashion to the {\nht} fits of \S\ref{snht} only with
the appropriate central frequencies and line strengths for the 7
hyperfine components~\citep{cas95,tin00} input into
Equation~\ref{taunht}. The excitation temperature from the fit and
temperature from \S\ref{sTk} then determine the density of the medium
using Equation~\ref{nH2}~\citep[see also][]{cas02}. The best fit to
the composite {\nthp} spectrum gives an average particle density
$n_{ex} = 4\times10^4$\,cm$^{-3}$ in {\core}. 

The average density inside this core using a mass of 2\,{\msun} and
volume of a prolate spheroid with dimensions from
Table~\ref{morphtable} is $5\times10^5$\,cm$^{-3}$. In consideration
of the factor of two error in the mass, the unknown 3-dimensional
geometry, and the estimates of $n_{ex}$ from the {\nht} and {\nthp}
data, the average density in {\core} $\langle n \rangle \approx
10^4$--$10^5$\,cm$^{-3}$.

\section{Discussion} \label{sdisc}
\subsection{Energy Partition} \label{senergy}
The gravitational stability and boundedness of {\core} can be
estimated through investigation of the energy partition.

The gravitational self-energy for an ellipsoidal mass with a
power-law density distribution, $\rho(r) \propto r^{-p}$,
is
\begin{equation}
E_{grav} = -\alpha_G\frac{3}{5}\frac{G\,M^2}{R}
\label{egrav}
\end{equation}
where
\begin{equation}
\alpha_G = \frac{(1-p/3)}{(1-2p/5)}\frac{\arcsin e}{e}
\label{alphaG}
\end{equation}
and $e$ is the eccentricity of the ellipsoid~\citep{ber92}.
An estimate of the rotational energy can be derived by assuming a
spherical geometry with the radius equal to the semi-major axis. 
\begin{equation}
E_{rot} \approx
\alpha_I\frac{2}{5}M\,R_{maj}^2\frac{\mathscr{G}^2}{\sin^2i} 
\label{erot}
\end{equation}
where $\mathscr{G}$ is the global velocity gradient
and $\alpha_I = (1-p/3)/(1-p/5)$. The thermal energy for a diatomic
molecular gas is $5/2NkT_k$ and the turbulent energy is estimated from
the non-thermal component of the line width $E_{turb} =
3/2M\sigma_{nt}^2$, where $\sigma_{nt} = 0.425 \Delta v_{nt}$.

Table~\ref{energytable} summarizes the results for the energy
partition in {\core} for three different power law density profiles, $p = 0.0$,
1.2, and 2.0, using both a spherical, $y = 1$, and prolate spheroidal
geometry with $y = 2$. In this table, $E_{vir} = 2\mathscr{T} +
\mathscr{W}$ and represents a measure 
of the gravitational stability based on virial arguments in the
absence of magnetic fields~\citep[{\eg}][]{mck92}. In this definition
$\mathscr{T} = E_{therm}+E_{turb}+E_{rot}$ is the kinetic
energy term in the absence of pressure confinement and $\mathscr{W} =
E_{grav}$ is the gravitational term. The negative values of $E_{vir}$
in Table~\ref{energytable} for all cases is strong evidence that
{\core} is confined by its own self-gravity. The values most
consistent with our analysis in \S\ref{sresults} are $p=1.2$ and
$y=2$.

The strength of the magnetic field is not known in this
region. Given the total energy estimates in Table~\ref{energytable}, a 
field strength of $B \gtrsim 160\,\mu$G is needed for {\core} to be
magnetically sub-critical which seems high based on Zeeman
measurements~\citep{lev01,cru03}, but is on the order of recent estimates
from polarized dust emission in starless cores~\citep{cru04}. The
ambient magnetic field position angle in L1551 was measured to be roughly 
80$^\circ$~\citep{vrb76} based on the polarization of background stars
in the periphery of the cloud. This angle is roughly consistent with the
orientation of the jets and outflows and may reflect a
relationship between the magnetic field geometry and the stellar
formation process in L1551. The minor axis of {\core} is 
also in rough alignment with the ambient magnetic field position
angle, but further study of the magnetic field in the immediate
vicinity of {\core} is needed to determine conclusively its
effects on the evolution of this molecular core. 

To infer the force balance of a core from energy considerations the
equilibrium state must be assumed. From \S\ref{smass}, $M_{mol}
\approx M_{vir}$ suggesting that {\core} is virialized. While virial
equilibrium is generally conjectured for ammonia cores~\citep[{\textsl
    e.g.}][]{jij99}, recent numerical results suggest that cores are
not likely to be hydrostatic and that dynamic cores projected onto the
sky can appear in equilibrium~\citep[][and references
  therein]{bal03}. Additionally, there are no equilibrium conditions
for an isolated, self-gravitating, prolate
spheroid~\citep{mye91}. Nevertheless, we can conclude from these
analyses that {\core} is gravitationally bound and may be
unstable. Thermal pressures provide some support against collapse, but
ultimately {\core} would have to be magnetically sub-critical to
prevent collapse on a free-fall time scale, $t_{ff} \approx 0.3$\,Myr.

\subsection{Core Environment}\label{senvi}
{\core} appears fairly simple traced solely in {\nht}, but is
significantly more complex when viewed in other molecular
tracers. Figure~\ref{spectra} shows the comparison between the
molecular emission of several tracers at Point A with the vertical
dotted line signifying the central LSR
velocity of the {\nht} (1,1) transition from
Table~\ref{morphtable}. All spectra in Figure~\ref{spectra} are
taken in the 3 mm band using the Kitt Peak 12~m Telescope except the {\cts}
spectrum which was taken from the 
BIMA 9 mm map, smoothed to $\sim 55${\arcsec} to minimize beam
dilution effects. 

The {\thco} spectrum at the top of Figure~\ref{spectra} is optically
thick; $\tau_{13} = 1.9 \pm 0.2$ using Equation~\ref{tauratio} with
the error determined by Monte Carlo simulation assuming no error in
the isotopic ratio. The width of this line is much larger than
expected from the direct conversion from the {\nht} width and there is
a dip in the line intensity at the central {\nht} velocity. The
{\thco} is likely tracing a column of gas which extends from the high
density interior into the more tenuous and turbulent surroundings.

The CS line in Figure~\ref{spectra} clearly shows a non-Gaussian shape
which represents all of our CS spectra taken across the major axis of
the {\cts} emission. This profile shape has a high peak at $v_{LSR}
\approx 6.5$\,{\kms} and a shoulder or second peak at $v_{LSR} \approx
7.0$\,{\kms} which could be due to another component along the line of
sight or self-absorption. There is a $\sim 50\%$ decrement in
intensity at the central velocity of the {\nht} emission, suggesting
that this feature is due to self-absorption. The broad width of the CS
line also means that this gas comprises a larger volume along the line
of sight than the {\nht}. It is possible that the asymmetric profile
is due to gas infall from a layer~\citep{mye96} or spherical
shell~\citep{eva99}.

Another intriguing feature seen in Figure~\ref{spectra} is the
redshifted peak of the {\cts} emission in relation to the central
velocity of the {\nht}. This 0.16\,{\kms} shift must be due to the
kinematics of the gas in {\core} since the transition frequency used
for our observations has an error~\citep{yam90} much less than the
observed line width. If the {\cts} emission lies on the near side of
{\core}, this implies that the {\cts} is moving toward the highest
density environs and may be further evidence of infall. Conversely,
this kinematic shift could be due to outflow from {\core}. Chemical
differentiation within molecular cores is a well known
phenomenon~\citep[][and references therein]{taf04} and could also be
contributing to the discrepancies seen between {\cts} and the other
molecular tracers. However, there is no clear evidence of depletion in
either {\thco} or {\ceo}.  Alternatively, the {\cts} could be tracing
a distinct component along the line of sight. Given the suggestive
profile of the CS($2-1$) line, the subsonic velocity shift of the
{\cts}, and the coincidence of the {\cts} peak with the optically
thick {\thco}, the interpretation that the {\cts} gas is in a state of
infall seems most likely.  

The {\nthp} spectrum in Figure~\ref{spectra} shows the isolated
$J_{F1,F} = 1_{0,1} \rightarrow 0_{1,2}$ component at Point
A. The fit to the full spectrum (see \S\ref{smass})
yields a value of $\tau_T = 5.5\pm0.3$ which means the optical depth
of the isolated component is  $\sim 0.6$. Assuming a beam filling
factor of unity, the excitation temperature $T_x = 4.0\pm0.1$\,K, and
the total column of {\nthp}
within the 12\,m beam is $5\pm0.5\times10^{12}$\,cm$^{-2}$. The beam
corrected column depth ratio of our two nitrogenated molecular
tracers, $N$({\nht})/$N(${\nthp})$ \approx 100$, is consistent with
other studies of starless cores in Taurus~\citep{ben98}. The
distribution of {\nthp} emission peaks slightly to the red of the
{\nht} and is about a factor of 2 broader than that expected from 9\,K
gas. The {\nthp} molecular emission is likely tracing a larger column
of gas than the {\nht}.

From these spectra, a picture emerges in which the cold,
gravitationally bound inner region of {\core} is surrounded by a
warmer, dynamic envelope showing evidence of gaseous infall. This
envelope may not be uniform, but is connected to the more static and
smoothly distributed, tenuous molecular gas pervading L1551.

\section{Conclusions} \label{sconclusions}
In this article, the diagnosis of a newly discovered molecular core
found within the L1551 dark cloud is described. This
core, referred to as {\core}, is located at $4^{\rm h}31^{\rm
  m}09^{\rm s}\!.9$,
$+18^\circ12^\prime41^{\prime\prime}\,\,(J2000)$ and has a full width
at half-maximum in {\nht} emission of $\sim 2^\prime\times 1^\prime$
at a position angle of 133$^\circ$ east of 
north. It has an inner region with a relatively flat density profile
which turns over to a more 
steeply declining function at a radius of $\sim 0.04$\,pc. {\core}
has a mass of $\sim 2$\,{\msun}, an average density of
$10^4$--$10^5$\,cm$^{-3}$, and a kinetic temperature of 9\,K which varies
by less than 50\% in the inner region. The
velocity of the turbulence is on the order of the sound
speed in the medium and comprises
4\% of the gravitational energy, $E_{grav}$, of the core. The
rotational energy is $\sim 
2.5\times10^{-3}\sin^{-2}i\,|E_{grav}|$, while the thermal energy is
0.3\,$|E_{grav}|$. The core is gravitationally bound and either will
collapse on a time scale of 0.3\,Myr or is supported by $\sim
160\,\mu$G magnetic fields. {\core} has a high density interior traced
by {\nht} and {\nthp} suspended within a dynamic, perhaps infalling
envelope which is connected to the more static and tenuous molecular
gas seen throughout L1551.

From energy considerations it appears as though {\core} will contract
to form a star or stellar system. However, there is future work that
could further characterize {\core} and more accurately place it within
the larger star forming context of L1551. An accurate measurement of
the magnetic field in {\core} will further our understanding of the
gravitational support mechanism and constrain the core lifetime. The
energy partition depends critically on the total mass and another
independent method of determining the total mass, such as from dust
emission, should be performed. While the state of the dense, inner
regions of {\core} are presented herein, supplementary information
regarding the envelope of this core is needed. Spectral line maps of
CS and its optically thin isotopomer, C$^{34}$S, will determine the
nature of the asymmetric CS line profile and provide a more complete
physical, spatial, and chemical picture of the envelope. Lastly, HH265
lies within the projected molecular emission of {\core} with no known
driving source. Further investigation into the possible relationship
between HH265 and {\core} may give deeper insight into the
evolutionary state of {\core}.

\acknowledgments
We thank Leo Blitz for allocating time for this project during the
BIMA 1~cm observing season, and John Carlstrom for donating the usage
of his 30\,GHz receivers. J. J. S. would like to thank Carl Heiles for
thoughtful input.This research has made use of the SIMBAD
database, operated at CDS, Strasbourg, France. This research was
partially supported by the National Science Foundation under grant
AST 02-28963.
Facilities: \facility{Berkeley-Illinois-Maryland-Association Array},
  \facility{Green Bank 100~m Telescope}, \facility{Kitt Peak 12~m
  Telescope}

\appendix

\section{Ammonia Emission Model} \label{snht}
We model our observed spectra using 
\begin{equation}
T_A^*(\nu) = A\left(1-e^{-\tau_\nu}\right)
\label{TA}
\end{equation}
where
\begin{equation}
A = \eta_B\,\eta_f\left[J_\nu(T_{ex}) - J_\nu(T_{bg})\right]
\label{A}
\end{equation}
and
\begin{equation}
J_\nu(T) = \frac{h\nu}{k}\frac{1}{\exp(h\nu/kT) - 1}
\label{JT}
\end{equation}
For the (1,1) transition, the optical depth is 
\begin{equation}
\tau_\nu = \tau_T\sum_{j=1}^{18}s_j\,
\exp\left[\frac{(\nu-\nu_j)^2}{2\sigma_\nu^2}\right]
\label{taunht}
\end{equation}
where $\tau_T$ is the total optical depth of the transition, $s_j$
are the relative line strengths of the 18 hyperfine components,
$\sigma_\nu$ is the standard deviation in frequency, and 
$\nu_j$ are the central frequencies of the hyperfine
transitions~\citep{kuk67,ryd77}. To fit the (1,1) transition we wrote
a wrapper for the IDL non-linear least-squares fitting routine
MPFIT\footnote{\url{http://astrog.physics.wisc.edu/$\sim$craigm/idl/fitting.html}}
which easily allows for the tying together of parameters and returns
all relevant fitting quantities such as the covariance matrix. 
More than 70\% of the total (2,2) intensity lies in the main component
for which there are 3 hyperfine transitions ($\Delta F_1 = 0$ and
$\Delta F = 0$) with separations $\sim
1$\,kHz~\citep{kuk67}. Therefore, a single Gaussian component is used
to fit our (2,2) spectra. 

Our non-linear least-squares fitting routine utilizes Equations~\ref{TA}
and~\ref{taunht} to fit our (1,1) spectra with the amplitude factor,
$A$; the LSR velocity of the emitting gas; the velocity width; and the
total optical depth as free parameters. The velocity width is then
analytically corrected for optical depth~\citep[{\eg}][]{phi79} and
instrumental broadening. The assumption that the excitation
temperatures for all hyperfine transitions are equal is inherent in
this fitting procedure.  

The relative population of {\nht} molecules in
$J=2$ and $J=1$ states can be solved for assuming equal excitation
temperatures and line widths for the (1,1) and (2,2)
transitions~\citep[\textsl{c.f.}][]{ho79}
\begin{equation}
T_R(2,1) = -T_0\left[\ln\left\{\frac{-g_{11}}{g_{22}} \frac{A_{11}}{A_{22}}
\left( \frac{\nu_{22}}{\nu_{11}} \right)^2 \frac{1}{\tau_{11}\,s_{22}}
\cdot \ln\left[1 - \frac{T_B(2,2)}{T_B(1,1)}
\left(1-e^{-\tau_{11}s_{11}}\right) \right]\right\}\right]^{-1}
\label{trot}
\end{equation}
where $T_0$ is the energy difference between the $J=1$ and $J=2$
states in units of Kelvin, $g_{11}$ and $g_{22}$ are the statistical
weights, $\tau_{11}$ is the total optical depth of the (1,1)
transition, and  $s_{11}$ and $s_{22}$ are the relative line strengths
of the transitions. 

The kinetic temperature, $T_k$, can be very well approximated from
$T_R$ if $T_k \ll T_0$ since transitions between K-ladders can only
happen via collision and non-metastable states within a K-ladder
radiatively decay very quickly~\citep{ho83,wal83,stu85}. Then, only
the (1,1), (2,2) and (2,1) states and their respective radiative and 
collisional rates need to be considered in the detailed balance
equation. Using~\cite{ho83} for the spontaneous emission rate
and~\cite{dan88} for the collisional rates, we obtain a transcendental
equation for $T_k$ in terms of $T_R$ and $T_0$
\begin{equation}
T_R = T_k\cdot
\left\{1+\frac{T_k}{T_0}\ln\left[1+0.6\exp\left(-15.7/T_k\right)\right]
\right\}^{-1}
\label{tk}
\end{equation}

If a specific value of the beam filling factor is assumed, the
excitation temperature can be obtained from the amplitude factor in
our fits. The volume particle density can then be estimated using the
equation of detailed balance for a 2-level system~\citep{stu85,cas02}. 
\begin{equation}
\frac{n_{ex}}{n^\prime_{crit}} = \frac{\tilde{T}_x -
  \tilde{T}_{bg}}{\left(h\nu/k\right)(1-\tilde{T}_x/\tilde{T}_k)}
\label{nH2}
\end{equation}
where
\begin{equation}
n^\prime_{crit}= \beta\cdot\frac{A_{ul}}{C_{ul}}.
\label{ncrit}
\end{equation}
$A_{ul}$ and $C_{ul}$ are the
spontaneous emission and collisional de-excitation rates for the
transition with $u$ and $l$ corresponding to the upper and lower
energy levels within the (1,1) state. The escape probability, $\beta$
should be obtained by detailed modeling, but here we use the
approximation,
\begin{equation}
\beta = \frac{1-e^{-\tau}}{\tau}
\label{beta}
\end{equation}
where $\tau$ is taken to be the maximum optical depth from
Equation~\ref{taunht}. In Equation~\ref{nH2},
\begin{equation}
\tilde{T} = \left(\frac{h\nu}{k}\right)
\left[1-\exp\left(\frac{-h\nu}{kT}\right)\right]^{-1}   
\label{tilde}
\end{equation}
$T_k$ is the kinetic temperature of the medium,
$T_{bg} = 2.73$\,K is the background radiation temperature, and
$n_{ex}$ is the particle density of the collisionally exciting
species, taken to be H$_2$.

\clearpage
\onecolumn
\begin{figure}
\centering
\includegraphics[angle=270,width=6in]{overview.ps}
\caption{In the right panel, the velocity integrated {\thco} in L1551
  is overlaid with pre-main sequence stars (stars), an incomplete
  sample of Herbig-Haro objects (triangles), known jet and
  outflow directions (arrows)~\citep{mun90,dev99}, and the spatial
  coverage of our {\nht} map (dashed box). {\cts} emission is revealed
  to the northwest of center in a relatively quiescent region of
  L1551. In the right hand panel, a close-up view of the {\cts}
  emission region is shown overlaid on the {\thco} map which has been
  restrectched to the maximum and minimum values in the region. 
  \label{overview}}
\end{figure}

\begin{figure}
\centering
\includegraphics[angle=270,width=6in]{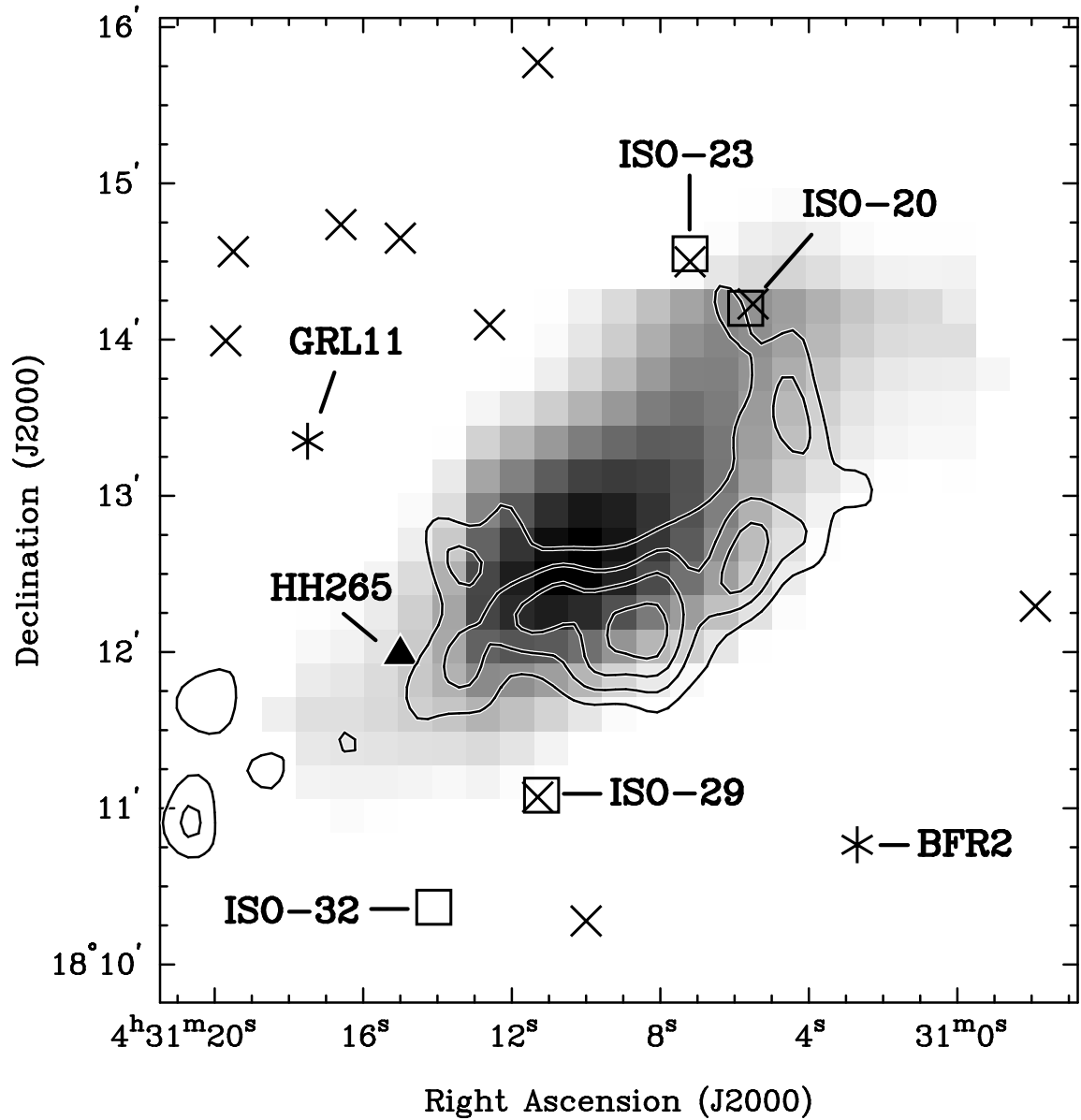}
\caption{Velocity integrated {\nht} emission in greyscale with {\cts}
  contours at 30, 50, 70, and 90\% the 0.623\,K\,{\kms} maximum
  overlaid. Boxes signify the point sources from~\cite{gal04},
  crosses are points sources from the 2MASS catalog, and the asterisks
  are SIMBAD objects. 
  \label{pointsource}}
\end{figure}

\begin{figure}
\centering
\includegraphics[angle=270,width=6in]{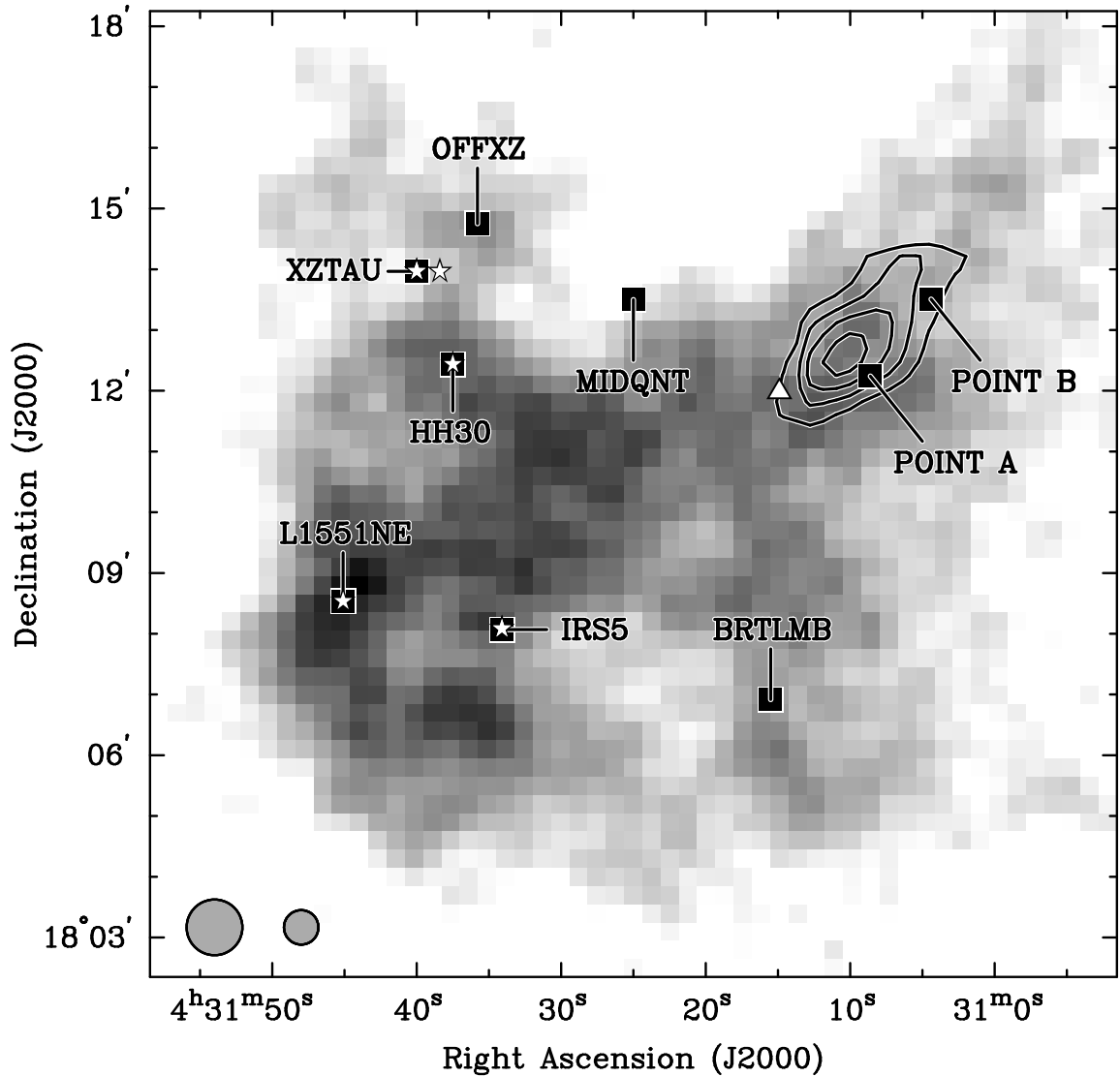}
\caption{Velocity integrated {\ceo} in L1551 with contours of
  {\nht}(1,1) overlaid at intervals of 30, 50 70 and 90\% the peak
  flux of 2.96\,K\,{\kms}. The labeled squares indicate the pointing
  centers of position switched 23\,GHz spectra, the white star symbols
  are the pre-main sequence stars of Figure~\ref{overview}, and the
  white triangle is the position of HH265. The Kitt Peak and GBT beams
  are shown from the bottom left respectively. \label{nh3plot}} 
\end{figure}

\begin{figure}
\centering
\includegraphics[angle=0,height=7.75in]{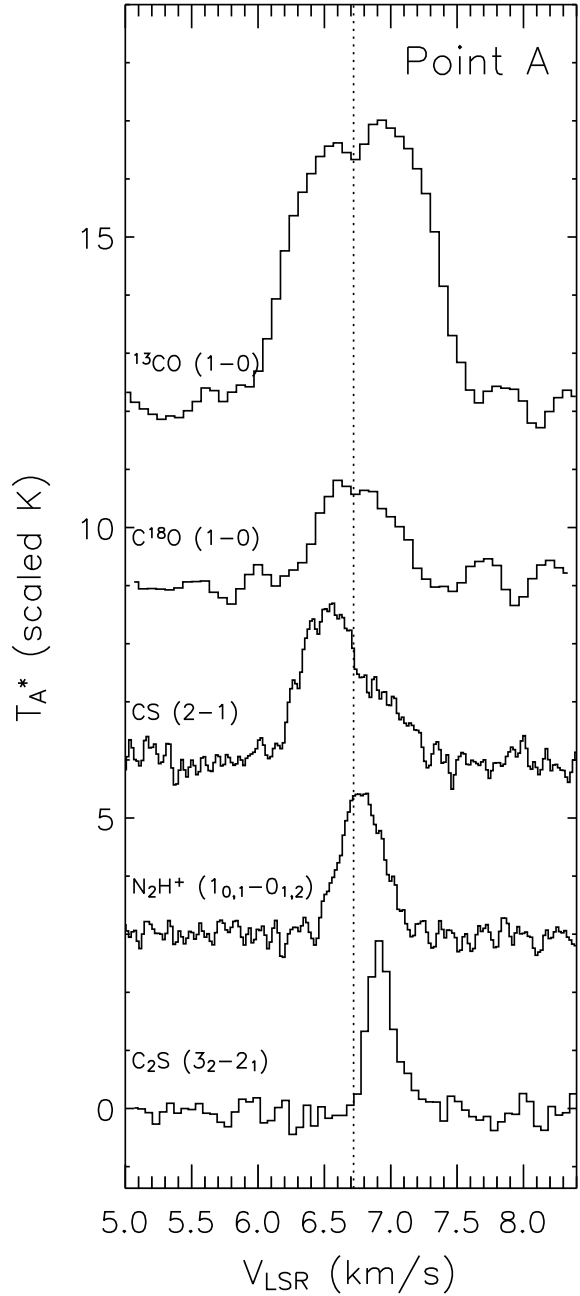}
\caption{Spectra from an assortment of molecular transitions at
  Point A. The CS and {\nthp} spectra are scaled by
  $5\times$ and the {\cts} spectrum is scaled by $2\times$. The vertical
  dotted line signifies the systemic velocity of the {\nht}
  emission at Point A. \label{spectra}} 
\end{figure}

\begin{figure}
\centering
\includegraphics[angle=270,width=6in]{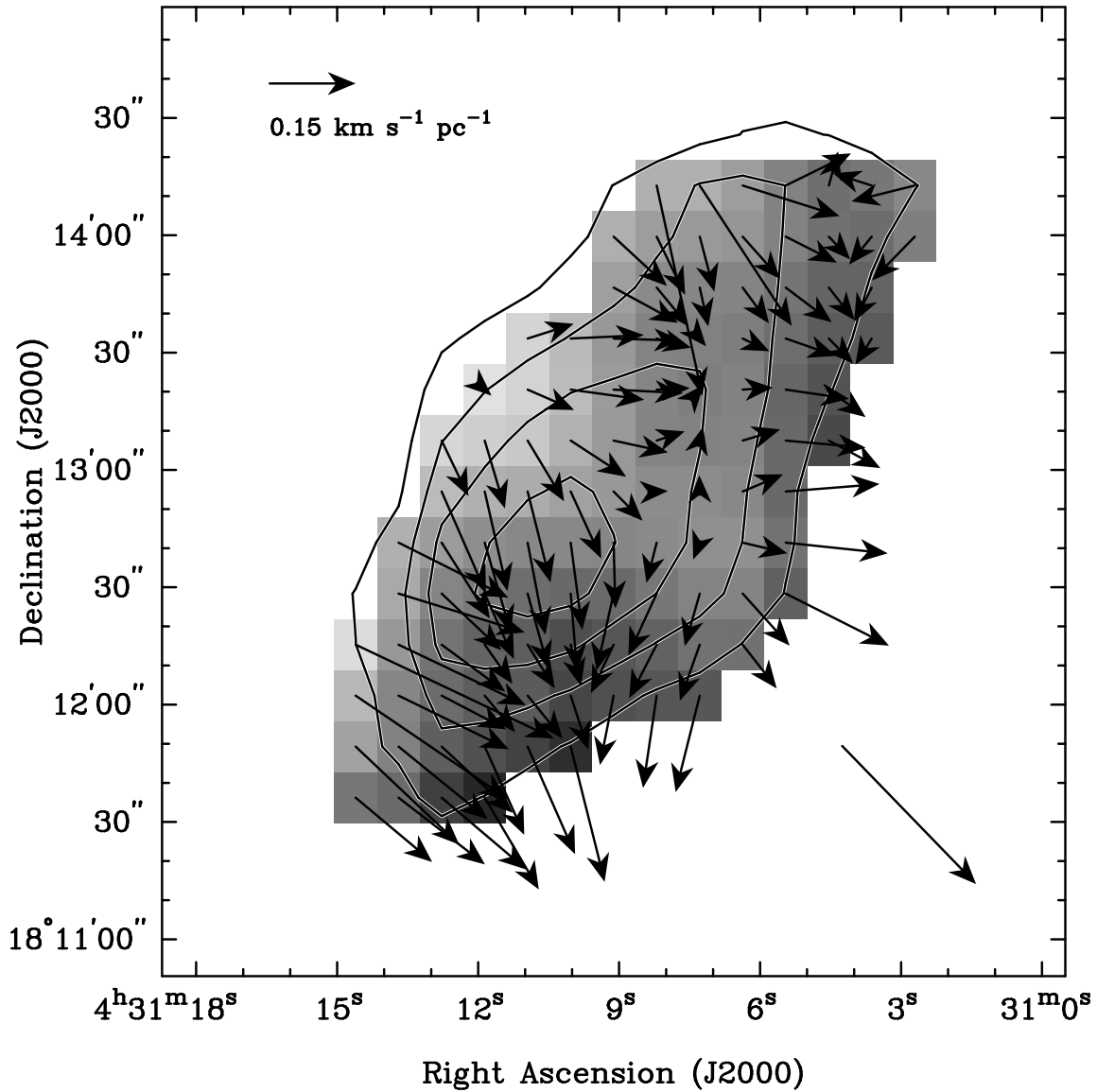}
\caption{First moment of {\nht} emission in greyscale is overlaid
  with local velocity gradients. Arrow lengths signify the relative
  gradient strengths. The 224$^\circ$ position angle of the global
  1.24\,{\kms}pc$^{-1}$ gradient is shown in the bottom right (length
  not to scale). Velocity integrated {\nht} intensity
  of the optically thin satellite components is shown in contour at
  30, 50, 70 and 90\% maximum intensity. \label{moment1}}
\end{figure}

\begin{figure}
\centering
\includegraphics[angle=90,width=6in]{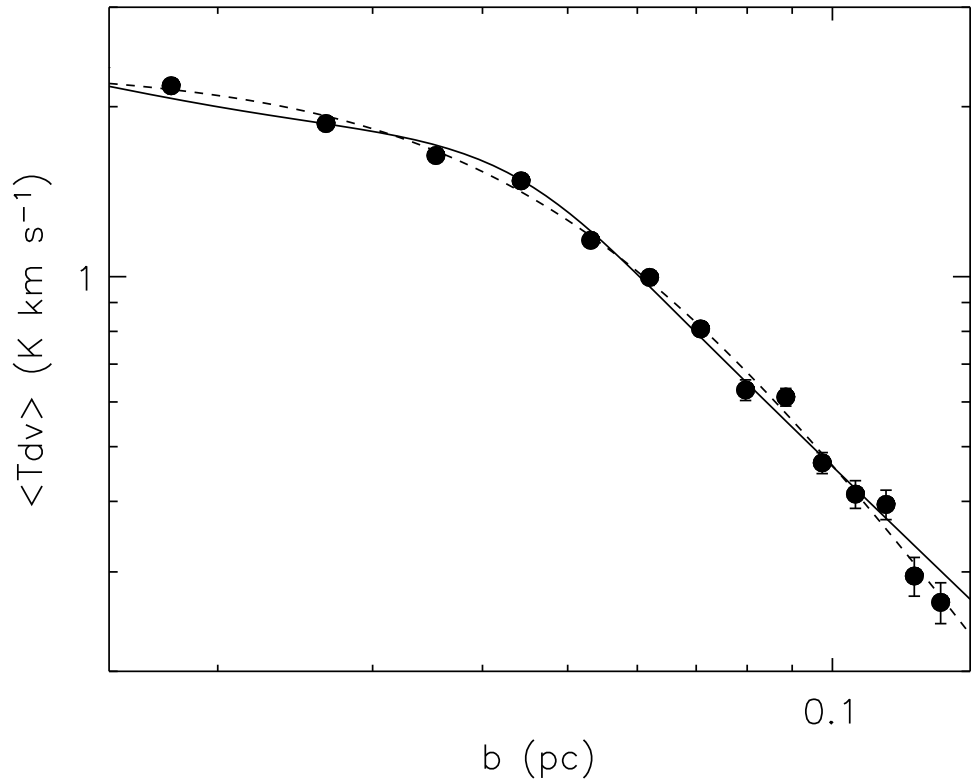}
\caption{Integrated intensity in the quadrupole satellite components
  of the {\nht}(1,1) transition as a function of impact
  parameter. Error bars are the root mean square deviation of the mean 
  intensity within an elliptical annulus and lie mostly within the
  filled circles. Chi-squared is minimized for models derived from
  Equations~\ref{brokenpl} and \ref{profilefunc} and the results are
  plotted as the solid and dashed lines respectively. \label{profile}}
\end{figure}

\begin{figure}
\centering
\includegraphics[angle=90,width=6.5in]{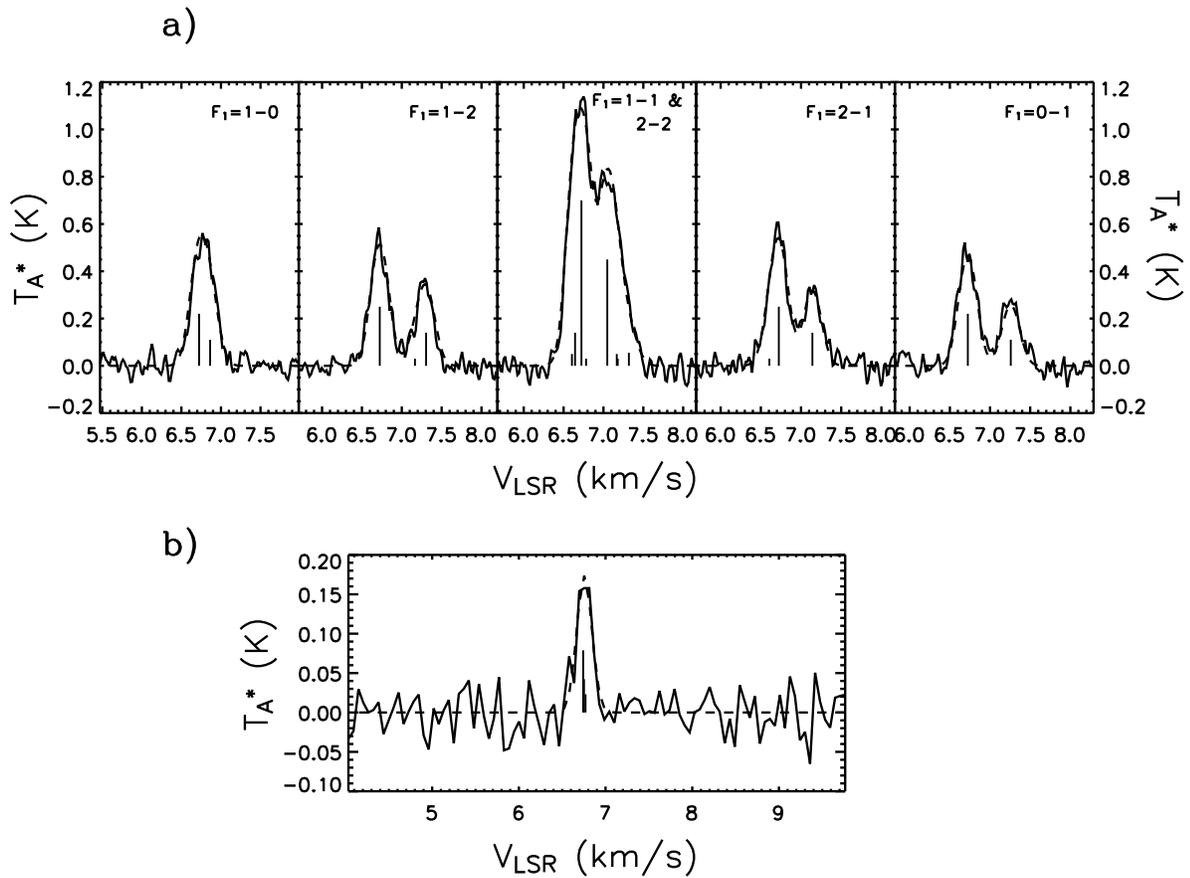}
\caption{(a) {\nht}(1,1) spectrum at Point A with \S\ref{snht} fit
  overlaid as the dashed line. The relative line strengths of
  all 18 hyperfine components are depicted as vertical lines and the
  LSR velocity scale in each quadrupole window is in reference to the
  strongest hyperfine line within. (b) The {\nht}(2,2) spectrum of 
  Point A with single Gaussian fit shown as the dashed line. Relative
  hyperfine line strengths are shown in the main
  component. \label{pointaspec}} 
\end{figure}

\clearpage

\begin{deluxetable}{llccc} 
\tablecolumns{5} 
\tablewidth{0pc} 
\tablecaption{Summary of Molecular Line Data\label{obstable}} 
\tablehead{
\multicolumn{5}{c}{BIMA} \\
\cline{1-5} \\
\colhead{Transitions} &
\colhead{Frequency (MHz)} & 
\colhead{$\Delta\nu$\tablenotemark{a} (kHz)} &
\colhead{Obs. Type} &
\colhead{$\theta_{beam}$} }
\startdata 
{\cts}($3_2 \rightarrow 2_1$) & 33751.3737\tablenotemark{1} &
6.1 & Mosaic & $27^{\prime\prime}\times20^{\prime\prime}$\\
{\thco}($1 \rightarrow 0$) & 110201.3541\tablenotemark{1} & 
24.4 & Mosaic & $10^{\prime\prime}\times9^{\prime\prime}$ \\ 
{\ceo}($1 \rightarrow 0$) & 109782.1734\tablenotemark{1} &
24.4 & Mosaic & $10^{\prime\prime}\times9^{\prime\prime}$ \\
CS($2 \rightarrow 1$) & 97980.9500\tablenotemark{1} & 
12.2 & Mosaic & $9^{\prime\prime}\times7^{\prime\prime}$ \\
{\nthp}($1 \rightarrow 0$) & 93176.2650\tablenotemark{2} & 
12.2 & Mosaic & $11^{\prime\prime}\times10^{\prime\prime}$  \\
{\ctfs}($2 \rightarrow 1$) & 109782.1734 &
12.2 & Mosaic & $11^{\prime\prime}\times10^{\prime\prime}$ \\
\cutinhead{Green Bank 100m}
{\nht}(1,1) & 23694.4955\tablenotemark{3,4} & 3.1 &
OTF\tablenotemark{b} & 34{\arcsec} \\
{\nht}(1,1) & 23694.4955 & 3.1 & APS\tablenotemark{c} & 34{\arcsec} \\  
{\nht}(2,2) & 23722.6333\tablenotemark{3} & 3.1 & OTF & 34{\arcsec} 
\\
{\nht}(2,2) & 23722.6333 & 3.1 & APS & 34{\arcsec} \\ 
\cutinhead{Kitt Peak 12m}
{\thco}($1 \rightarrow 0$) & 110201.3541 & 24.4 & OTF & 55{\arcsec} \\
{\ceo}($1 \rightarrow 0$) & 109782.1734 & 24.4 & OTF & 55{\arcsec} \\
{\nthp}($1 \rightarrow 0$) & 93176.2650\tablenotemark{2} & 12.2 &
FS\tablenotemark{d} & 65{\arcsec} \\
CS($2 \rightarrow 1$) & 97980.9500 & 12.2 & FS &
65{\arcsec} \\ 
\enddata  
\tablenotetext{a}{Post-reduction spectral resolution}
\tablenotetext{b}{On-the-Fly mapping}
\tablenotetext{c}{Absolute Position Switching}
\tablenotetext{d}{Frequency Switching}
\tablenotetext{1}{From \cite{pic98}}
\tablenotetext{2}{Frequency of the $J_{F1,F} =
  1_{0,1}~\rightarrow~0_{1,2}$ hyperfine component~\citep{cas95}} 
\tablenotetext{3}{Hyperfine weighted mean transition frequency
  from~\cite{kuk67}}
\tablenotetext{4}{\cite{ryd77}}
\end{deluxetable} 

\clearpage

\begin{deluxetable}{rccccc} 
\tablecolumns{6} 
\tablewidth{0pc} 
\tablecaption{Summary of the GBT 23\,GHz Position Switched
  Observations\label{nh3pnts}} 
\tablehead{
\colhead{Source} & \colhead{R.A.(J2000)} &
\colhead{Dec.(J2000)} & 
\colhead{T$_B$(1,1)\tablenotemark{a}} & 
\colhead{T$_B$(2,2)\tablenotemark{a}} & 
\colhead{$\sigma_{T_B}$}}
\startdata 
POINT A & 4:31:08.6 & +18:12:14.4 &   1.41  &  0.23   & 0.04 \\
POINT B & 4:31:04.4 & +18:13:30.4 &   1.09  & \nodata & 0.07 \\
IRS5    & 4:31:34.1 & +18:08:04.1 &   2.81  &  0.53   & 0.10 \\
L1551NE & 4:31:45.1 & +18:08:32.0 &   0.44  & \nodata & 0.06 \\  
XZTAU   & 4:31:40.0 & +18:13:58.0 & \nodata & \nodata & 0.06 \\
HH30    & 4:31:37.5 & +18:12:26.0 & \nodata & \nodata & 0.06 \\
OFFXZ   & 4:31:35.8 & +18:14:45.0 & \nodata & \nodata & 0.06 \\
MIDQNT  & 4:31:25.0 & +18:13:30.0 & \nodata & \nodata & 0.06 \\
BRTLMB  & 4:31:15.5 & +18:06:55.0 & \nodata & \nodata & 0.06 \\
\enddata 
\tablecomments{Absence of data signifies a non-detection.}
\tablenotetext{a}{Peak value of main component brightness temperature
  in Kelvin. Beam filling factors are assumed to be unity. }
\end{deluxetable}

\clearpage 

\begin{deluxetable}{ccccccc} 
  \tablecolumns{7} 
  \tablewidth{0pc}
  \tablecaption{Properties of {\nht} Emission in
  {\core}\label{morphtable}} 
  \tablehead{
    \colhead{R.A.\tablenotemark{a}} & \colhead{Dec.\tablenotemark{a}} &
    \colhead{$v_{LSR}$\tablenotemark{b}} & 
    \colhead{$\Delta v$\tablenotemark{b}} & 
    \colhead{maj.\tablenotemark{c}} & 
    \colhead{min.} & 
    \colhead{P.A.} \\ 
    \colhead{(J2000)} & \colhead{(J2000)} &
    \colhead{({\kms})} & 
    \colhead{({\kms})} & 
    \colhead{(arcmin)} & 
    \colhead{(arcmin)} & 
    \colhead{degrees} }
  \startdata 
   4:31:09.9 & +18:12:41 & 6.72 & 0.26 & 2.25 & 1.11 & 133 \\
  \enddata 
  \tablenotetext{a}{Coordinates of the peak emission.}
  \tablenotetext{b}{The line profile characteristics are derived from the mean
  spectrum averaged within a half-maximum contour of the {\nht}
  emission.}
  \tablenotetext{c}{``maj.'' and ``min.'' are the FWHM of the major and
  minor axes respectively.}
\end{deluxetable} 

\clearpage

\begin{deluxetable}{rccrccccrcrr} 
  \rotate
  \tablecolumns{12} 
  \tablewidth{0pc}
  \tablecaption{{\nht} (1,1) and (2,2) Fit Results \label{nh3qtable}}
  \tablehead{
    \colhead{} & \colhead{} & \colhead{} & \colhead{} & \colhead{} &
    \colhead{} & 
    \multicolumn{3}{c}{$\eta_f = 1$} & \colhead{} &
    \multicolumn{2}{c}{$T_x = T_k$} \\  
    \cline{7-9} \cline{11-12} \\
    \colhead{Source} & \colhead{$v_{\rm LSR}$} &
    \colhead{$\Delta v_{obs}$} &  
    \colhead{$T_k$\tablenotemark{a}} &
    \colhead{$\tau_T$} & 
    \colhead{} & \colhead{$T_x$} & 
    \colhead{$N$({\nht})} & \colhead{$n_{ex}$\tablenotemark{b}} & \colhead{} &
    \colhead{$N$({\nht})} & \colhead{$\eta_f$} \\
    \colhead{} &
    \colhead{(\kms)} &
    \colhead{(\kms)} &
    \colhead{(K)} &
    \colhead{} &
    \colhead{} &
    \colhead{(K)} &
    \colhead{(cm$^{-2}$)} &
    \colhead{(cm$^{-3}$)} &
    \colhead{} &
    \colhead{(cm$^{-2}$)} &
    \colhead{} }
  \startdata 
  POINT A & 6.72 & 0.24 & 8.7 & 5.8 &  & 4.4 & $2.1\times10^{15}$ & 
  $5\times10^3$ & & $3.7\times10^{14}$ & 0.26  \\  
  POINT B & 6.71 & 0.23 & $<12.4$ & 2.9 & & 4.5 & $9.2\times10^{14}$ & 
  $>4\times10^3$ & & $>1.5\times10^{14}$ & $>0.18$ \\
  IRS5 & 6.42 & 0.48 & 10.0 & 3.5 & & 6.4 & $8.2\times10^{14}$ & 
  $2\times10^4$ & & $4.1\times10^{14}$ & 0.47  \\   
  L1551NE & 6.55 & 0.40 & $<20.6$ & 4.1 &  & 3.2 &$1.3\times10^{16}$&
  $>8\times10^2$ & & $>4.6\times10^{14}$ & $>0.02$ \\ 
  \enddata 
  \tablenotetext{a}{Upper limits to $T_k$ derived from $3\,\sigma$
    level in the (2,2) spectrum. Quantities dependent on $T_k$
  were computed with upper limit value when appropriate.}
  \tablenotetext{b}{Density derived using equation
  Equations~\ref{nH2}--\ref{beta}}
\end{deluxetable} 

\clearpage

\begin{deluxetable}{ccccccc} 
  \tablecolumns{7} 
  \tablewidth{0pc}
  \tablecaption{{\core} Physical
  Characteristics\tablenotemark{a}\label{phystable}} 
  \tablehead{
    \colhead{$T_k$} & 
    \colhead{$M_{mol}$} & 
    \colhead{$M_{vir}$} & 
    \colhead{$\langle n \rangle$} &
    \colhead{$\Delta v_{th}$} & 
    \colhead{$\Delta v_{nt}$} & 
    \colhead{$\Delta v_{tot}$} \\
    \colhead{(K)} & 
    \colhead{({\msun})} & 
    \colhead{({\msun})} & 
    \colhead{(cm$^{-3}$)} &
    \colhead{({\kms})} & 
    \colhead{({\kms})} & 
    \colhead{({\kms})} }
  \startdata 
  9 & 2 & 1.9 & $5\times10^4$ & 0.42 & 0.20 & 0.46 \\
  \enddata 
  \tablenotetext{a}{See text for a discussion of the uncertainties.}
\end{deluxetable} 

\clearpage

\begin{deluxetable}{ccccccccc} 
  \tablecolumns{9} 
  \tablewidth{0pc}
  \tablecaption{Energy Partition in L1551-MC\label{energytable}} 
  \tablehead{
    \colhead{$p$\tablenotemark{a}} & 
    \colhead{$y$\tablenotemark{b}} & 
    \colhead{$\alpha_G$} &
    \colhead{$\alpha_I$} &    
    \colhead{$E_{grav}$} & 
    \colhead{$E_{therm}$} &
    \colhead{$E_{turb}\times10^2$} & 
    \colhead{$E_{rot}$\tablenotemark{c}$\times10^3$} &
    \colhead{$E_{vir}$\tablenotemark{d}} \\ 
    \colhead{} & 
    \colhead{} & 
    \colhead{} & 
    \colhead{} & 
    \colhead{($10^{43}$ergs)} & 
    \colhead{($|E_{grav}|$)} &
    \colhead{($|E_{grav}|$)} & 
    \colhead{($|E_{grav}|$)} & 
    \colhead{($|E_{grav}|$)} } 
  \startdata 
  0.0 & 1.0 & 1.00 & 1.00 & -0.9 & 0.28 & 3.2 & 2.8 & -0.37 \\
  1.2 & 1.0 & 1.15 & 0.79 & -1.1 & 0.24 & 2.7 & 1.9 & -0.46 \\
  2.0 & 1.0 & 1.67 & 0.56 & -1.5 & 0.17 & 1.9 & 0.9 & -0.63 \\
  0.0 & 2.0 & 0.76 & 1.00 & -0.7 & 0.37 & 4.1 & 3.7 & -0.18 \\
  1.2 & 2.0 & 0.87 & 0.79 & -0.8 & 0.32 & 3.6 & 2.5 & -0.29 \\
  2.0 & 2.0 & 1.27 & 0.56 & -1.2 & 0.22 & 2.5 & 1.2 & -0.51 \\
  \enddata 
  \tablenotetext{a}{Power index for $\rho(r) \propto r^{-p}$.}
  \tablenotetext{b}{Axial ratio, $y = b/a$} 
  \tablenotetext{c}{An inclination angle, $i = 90^\circ$ is assumed}
  \tablenotetext{d}{$E_{vir} = 2\mathscr{T} + \mathscr{W}$}
\end{deluxetable}

\end{document}